\documentclass[12pt]{article} 
\usepackage{epsfig}
\usepackage{amsfonts}
\usepackage{latexsym}
\usepackage{amsmath,amssymb}
\usepackage{verbatim}
\usepackage{mathtools}
\usepackage[dvipdfm,hypertex]{hyperref}
\def\half{{1\over 2}}
\numberwithin{equation}{section}

 \def\p{\partial}

\def\k{\kappa}

\newcommand{\bea}{\begin{eqnarray}}
\newcommand{\eea}{\end{eqnarray}}
\newcommand{\be}{\begin{equation}}
\newcommand{\ee}{\end{equation}}
\newcommand{\ba}{\begin{align}}
\newcommand{\ea}{\end{align}}

\newcommand{\W}{\mathcal{W}}
\newcommand{\tr}{\mbox{tr}}

\newcommand{\hs}[1]{\mbox{hs$[#1]$}}
\newcommand{\w}[1]{\mbox{$\W_\infty[#1]$}}
\newcommand{\bif}[2]{\small\left(\!\!\begin{array}{c}#1 \\#2\end{array}\!\!\right)}

  \makeatletter
  \let\over=\@@over \let\overwithdelims=\@@overwithdelims
  \let\atop=\@@atop \let\atopwithdelims=\@@atopwithdelims
  \let\above=\@@above \let\abovewithdelims=\@@abovewithdelims

\begin{document}

\ \\
\vspace{0.8cm}

\begin{center}

{ \LARGE {\bf  Symmetries of Holographic Minimal Models}}

\vspace{1.2cm}

Matthias R. Gaberdiel$^{a,b}$ and Thomas Hartman$^b$

\vspace{0.9cm}

{\it $^a$Institut f\"ur Theoretische Physik, ETH Zurich, \\
CH-8093 Z\"urich, Switzerland \\
 }

\vspace{0.5cm}

{\it $^b$ School of Natural Sciences, Institute for Advanced Study,
\\
Princeton, NJ 08540, USA \\ }

\vspace{0.5cm}

\tt{\small gaberdiel@itp.phys.ethz.ch, hartman@ias.edu} 

\vspace{1.3cm}

\end{center}

\begin{abstract}
It was recently proposed that a large $N$ limit of a family of minimal model CFTs is
dual to a certain higher spin gravity theory in AdS$_3$, where the 't~Hooft coupling constant of the CFT 
is related to a deformation parameter of the higher spin algebra. We identify the asymptotic
symmetry algebra of the higher spin theory for generic 't~Hooft parameter, and show that 
it coincides with a family of $\W$-algebras previously discovered in the context of the KP hierarchy. We furthermore demonstrate that this family of  $\W$-algebras controls the
representation theory of the minimal model CFTs in the 't~Hooft  limit. This provides a non-trivial
consistency check of the proposal and explains part of the underlying mechanism.
\end{abstract}

\pagebreak
\setcounter{page}{1}

\tableofcontents

\section{Introduction}
Simplified versions of the AdS/CFT correspondence allow for a detailed study of holography that would 
be impossible in a full string theory setting.  A promising approach along these lines is the 
investigation of higher spin theories of gravity in anti de Sitter space, which include a large 
(possibly infinite) number of fields with spins $s=2,3,\dots,N$, including the graviton of spin $s=2$.  
The higher spin fields take the place of string excitations.  Holography in this context is, to some extent, 
a complicated but tractable field redefinition \cite{Koch:2010cy,Douglas:2010rc}. This puts holography (in some special cases) on a footing similar to, say, Coleman's sine-Gordon/Thirring duality \cite{Coleman:1974bu} where 
operators of the two theories have a known one-to-one map, rather than the more mysterious 
strong-weak dualities of string theory.

In four bulk dimensions, Klebanov and Polyakov \cite{Klebanov:2002ja} conjectured a duality between 
Vasiliev's higher spin theory \cite{Vasiliev:2003ev,Bekaert:2005vh} and the $O(N)$ vector model at its two 
isolated critical points.  Recently interest in higher spin dualities was renewed by detailed checks of the 
correlation functions \cite{Giombi:2009wh}.  The bulk computations are difficult (though they can be 
simplified by a gauge choice \cite{Giombi:2010vg}), partly because the action of the higher spin theory is 
unknown.

In three bulk dimensions, the higher spin theory is much simpler.  The massless sector is described 
semiclassically by the Chern-Simons action, and thus the graviton and its higher spin cousins have no 
propagating modes.  Furthermore, the dual CFT is two dimensional and therefore subject to the strong 
constraints of the Virasoro algebra and its higher spin analogs, the $\W$-algebras.  This allows for 
full control of the dual CFT for all values of the coupling $\lambda$, without supersymmetry.  

The prime examples of exactly solvable interacting CFTs in two dimensions are the 
Virasoro minimal models, with central charge $c <1$.  These theories, which include experimentally 
relevant systems such as the Ising model, are not dual to any semiclassical gravity-like theory in AdS$_3$ 
because they do not have enough degrees of freedom to account for the large Brown-Henneaux central 
charge \cite{Brown:1986nw} of AdS$_3$ gravity, $c = 3\ell/2G \gg 1$, where $\ell$ is the AdS radius and 
$G$ is Newton's constant.  In other words, they have small $N$.  However, the underlying Virasoro 
symmetry can be extended to a larger $\W_N$-symmetry with conserved currents of dimensions 
$s=2,\dots,N$.  (The pure Virasoro symmetry is then simply $\W_2$.) The corresponding generalization 
of the Virasoro minimal models, called the $\W_N$-minimal models, allow for a large-$N$ limit and 
therefore for a potential interpretation as gravity in AdS$_3$. Other solvable CFTs admitting large 
$N$ limits were discussed recently in \cite{Kiritsis:2010xc}.

Following the appearance of $\W$-symmetry in the asymptotic algebra of AdS$_3$ higher spin 
theories \cite{Henneaux:2010xg,Campoleoni:2010zq,Gaberdiel:2010ar}, it was proposed that a 
particular AdS$_3$ higher spin theory together with a pair of massive complex scalar fields
is dual to the $\W_N$ minimal models at large $N$ 
\cite{Gaberdiel:2010pz}. Just like the Virasoro minimal models, the $\W_N$ models are parametrized 
by a level $k$. The large $N$ limit of \cite{Gaberdiel:2010pz} is taken in such a way that the 
't~Hooft coupling,  defined by
\be
\lambda = {N\over N+k} \ , \quad 0 < \lambda < 1 
\ee
is held fixed. 
The dual bulk theory has an infinite tower of massless higher spin fields as well as two complex 
scalars with masses
\be
M^2 = \lambda^2-1 \ .
\ee
The partition functions of the two theories were compared in \cite{Gaberdiel:2010pz}, and the first 
few terms (as an expansion in the modular parameters $q$ and $\bar{q}$) were found to match 
precisely at arbitrary 't~Hooft coupling $\lambda$.  

This can be compared to the string theory realization of AdS$_3$/CFT$_2$ where the 
boundary theory is the D1-D5 CFT \cite{Strominger:1996sh,Maldacena:1997re}.  
This theory has a simple point in moduli space, the orbifold point, where many quantities can be 
computed explicitly. However to compare to the gravity theory, the results must be extrapolated to 
strong coupling, and thus only certain protected quantities can be expected to match.  
In our context of minimal model holography, the CFT can be solved for any value of the 
't~Hooft parameter, and can be directly compared to the bulk. 
We should note that, at least on the face of it, minimal
model holography is only expected to be exact in the large $N$ limit (possibly including 
$1/N$ corrections on both sides), but does not hold directly at finite $N$. 
For example, it is not clear if or how black holes are captured by the finite $N$ minimal 
model  \cite{Castro:2010ce}.
\medskip

In this paper we give further evidence for the minimal model holography of \cite{Gaberdiel:2010pz}. 
To this end we study the higher spin bulk theory whose massless sector is described
by the Chern-Simons theory based on the infinite-dimensional 
Lie algebra \hs{\lambda}. (\hs{\lambda} is the higher spin analogue of
sl$(2)$ which in turn is relevant for the description of pure gravity on AdS$_3$; the Chern-Simons
theory based on \hs{\lambda} describes the massless sector of the bulk theory in 
\cite{Gaberdiel:2010pz}.)  For general 
't~Hooft parameter $\lambda$ we analyze the asymptotic symmetry algebra of this bulk theory,
generalizing the analysis of \cite{Henneaux:2010xg} and \cite{Campoleoni:2010zq}  for 
$\lambda=\half$ and $\lambda=N$, respectively. 
We show that, for general $\lambda$, the resulting algebra agrees with (two copies of) a specific
$\W_\infty[\lambda]$ algebra that has been previously constructed in the context of integrable systems by 
Figueroa-O'Farrill, Mas, and Ramos \cite{FigueroaO'Farrill:1992cv} and independently by 
Khesin and Zakharevich \cite{Khesin:1993ru,Khesin:1993ww}. 
This $\W$-algebra has generating fields of dimension $s=2,3,\dots$, which extends 
\hs{\lambda} much like the Virasoro algebra extends sl$(2)$.  
The algebra is in general nonlinear, 
meaning the commutation relations involve polynomials in the generating fields; an explicit description is 
given in appendix  \ref{a:fmr}. 

The algebra $\W_\infty[\lambda]$ is in fact related to various $\W_\infty$-algebras that have appeared 
in the literature before.  When $\lambda = 1$, all nonlinearities can be removed by a change of basis, 
and the algebra becomes the well known linear $\W_\infty^{\rm PRS}$ algebra of Pope, 
Romans and Shen \cite{Pope:1989ew}.  When $\lambda=N$, the trace that appears in the 
Chern-Simons action degenerates and all fields of spins $s > N$ should be removed; the bulk theory 
then reduces to sl$(N)$ Chern-Simons theory and the boundary algebra becomes 
$\W_N$ \cite{Campoleoni:2010zq}.

Given the usual AdS/CFT dictionary, the $\W_\infty[\lambda]$ algebra should now control 
the spectrum of the dual CFT. The representation theory of $\W_\infty[\lambda]$ is largely determined 
by the representation theory of the  global part  \hs{\lambda} of the bulk symmetry, since 
\hs{\lambda} can be identified with the so-called `wedge algebra'  whose general construction
was explained in \cite{Bowcock:1991zk}. 
We give fairly non-trivial evidence that the large-$N$ 't~Hooft limit of the $\W_N$ minimal
model representations come indeed from representations of \hs{\lambda}. In particular, we show 
that the eigenvalues of the spin-3 zero mode agree (up to some overall normalization which is
ambiguous) on the simplest representations. Furthermore we demonstrate that the characters of these
$\W_\infty[\lambda]$ representations reproduce precisely the characters of the
corresponding $\W_N$ minimal model representations in the 't~Hooft limit whose
first few terms were determined in  \cite{Gaberdiel:2010pz}.

The results of this paper therefore explain part of the match found in  \cite{Gaberdiel:2010pz}. 
More importantly, the detailed understanding 
of the symmetries provides a framework to analyze the bulk/boundary map in more detail.  
The full partition functions compared term-by-term in \cite{Gaberdiel:2010pz} were organized 
into representations of sl$(2)$; rearranging the results to be manifestly \hs{\lambda}-invariant 
greatly simplifies the task of proving equality to all orders \cite{workinprog}.
\medskip

The paper is organized as follows.  In section \ref{s:review}, we review the higher spin algebras 
\hs{\lambda}, general properties of $\W$-algebras, and the identification of the higher-spin algebra 
as the global or `wedge' component of a $\W$-algebra; everything in section \ref{s:review} is review.  
In section \ref{s:asg}, we compute the asymptotic symmetries of the \hs{\lambda} higher spin theory, 
derive the resulting algebra \w{\lambda}, and describe the relation of \w{\lambda} to various 
$\W_\infty$-algebras that have appeared in the literature.  In section \ref{ss:CFT} we give evidence that the 
global symmetries of the boundary CFT are indeed \hs{\lambda}.  Some details about the structure of the 
\hs{\lambda} algebra are given in appendix \ref{a:structure}, and the full commutation relations of 
\w{\lambda} are spelled out in appendix \ref{a:fmr}.

\section{$\W$-symmetry and Higher Spin Algebras}\label{s:review}

Let us begin by reviewing higher spin algebras, $\W$-algebras, and the connection between the two. 

In ordinary AdS$_3$ gravity, the bulk isometries are the six generators of sl$(2)\;\oplus\; $sl$(2)$.
Near the conformal boundary, these are enhanced to the Virasoro$\; \oplus$Virasoro symmetries 
of the dual CFT.  Conversely, the exact symmetries of the bulk can be recovered from the CFT by 
starting with the Virasoro generators $L_n, \bar{L}_n$ and restricting to the global subalgebra 
$n=0,\pm 1$.

A similar relationship connects higher spin algebras to $\W$-algebras.  The analogue of the bulk
isometries 
sl$(2)$ is the higher spin algebra \hs{\lambda}. It is an infinite-dimensional Lie algebra that
has a simple description in terms of the universal enveloping algebra $U({\rm sl}(2))$, as
we shall review in Section~\ref{ss:hsalgebra} below. The symmetry algebra of the dual CFT ({\it i.e.}\ the 
analogue of the Virasoro algebra) is a $\W$-algebra that we shall denote by \w{\lambda}. 
For a range of values of $\lambda$, it can be understood as the 't~Hooft large $N$ limit of
the minimal model $\W_N$-algebras. For generic $\lambda$ it has non-linear commutation relations,
as we shall demonstrate by an explicit calculation in Section~\ref{s:asg}. Because of these non-linearities,
\hs{\lambda} is not a subalgebra of \w{\lambda}. However, even for non-linear $\W$-algebras, there exists 
a standard construction by means of which one can associate the `finite' or `global' wedge algebra 
to it, {\it i.e.}\ the analogue of sl$(2)$ for the case of Virasoro; this will be explained 
in Section~\ref{ss:wedge}. 

\subsection{Higher Spin Algebras \hs{\lambda}}\label{ss:hsalgebra}

The one-parameter family of higher spin Lie algebras \hs{\lambda} has generators
\be
V^s_n \ , \quad s\geq 2 \ , \quad |n|<s  \ .
\ee
$V^2_{0,\pm 1}$ forms an sl$(2)$ subalgebra under which $V^s_n$ has spin $s-1$, 
\be\label{twoaction}
[V^2_m, V^s_n] = (-n + m(s-1)) V^s_{m+n} \  .
\ee
(Bulk fields associated to $V^s_n$ will have spacetime spin $s$.)  The full commutation relations are
\be\label{vcom}
[V^s_m, V^t_n] = \sum_{ \stackrel{u=2}{\vspace{0.1cm} \mbox{\tiny even}}}^{s+t-1}  g_u^{st}(m,n;\lambda) V_{m+n}^{s+t-u}
\ee
with structure constants $g_u^{st}(m,n;\lambda)$ given in appendix \ref{a:structure} \cite{Pope:1989sr}.
\smallskip

For the following another description of \hs{\lambda} will be important 
 \cite{Feigin88,Bordemann:1989zi,Bergshoeff:1989ns,Pope:1989sr}.  Consider the quotient of the universal enveloping algebra $U({\rm sl}(2))$ by the ideal generated by $(C_2-\mu {\bf 1})$, 
\be\label{idealcon}
B[\mu] = {U(\mbox{sl}(2))\over \langle C_2 - \mu{\bf 1} \rangle} \ .
\ee
Here $C_2$ is the quadratic Casimir of sl$(2)$; if we denote the generators of 
sl$(2)$ by  $J_0, J_\pm$ with commutation relations
\be
{}[J_+,J_-] = 2 J_0 \ , \qquad [J_\pm,J_0] = \pm J_\pm  \ ,
\ee
then $C_2$ is given by 
\be\label{cas}
C_2 \equiv J_0^2 - \half(J_+ J_- + J_- J_+)  \ .
\ee
Unitary representations of sl$(2)$ have $C_2 > -\tfrac{1}{4}$, so we parameterize the Casimir as
\be
\mu = {1\over 4}(\lambda^2-1) \ .
\ee
The Lie algebra \hs{\lambda} can be identified (as a vector space) with a subspace of $B[\mu]$,
\be\label{idealcon1}
\hs{\lambda} \oplus \mathbb{C} = B[\mu] \ .
\ee
The vector corresponding to ${\mathbb C}$ in (\ref{idealcon1}) is the identity generator ${\bf 1}$
of the universal enveloping algebra, which one may formally identify with $V^1_0$. 
The modes $V^2_{0,\pm 1}$ in the sl$(2)$ subalgebra
of \hs{\lambda} can be identified with $J_{0,\pm 1}$, respectively, while for $n\geq 2$ 
\be\label{env}
V^s_n = (-1)^{s-1-n} \frac{(n+s-1)!}{(2s-2)!} \, 
 \Bigl[ \underbrace{J_- , \dots [J_-, [J_-}_{\hbox{\footnotesize{$s-1-n$ terms}}}, J_+^{s-1}]]\Bigr] \ .
\ee
The vector space $B[\mu]$ in (\ref{idealcon}) is an associative algebra whose product 
we denote by $\star$. The Lie algebra structure of $\hs{\lambda} \oplus {\mathbb C}$ is then defined by the 
commutator $[X,Y] = X \star Y - Y \star X$. Note that the identity generator ${\bf 1}$ is central. On $B[\mu]$
we can define an invariant bilinear trace \cite{Vasiliev:1989re} via
\be\label{quadform}
\tr(X \star Y) =  \left. X\star Y \right|_{J_a=0} \ ,
\ee
{\it i.e.}\ by retaining only the term proportional to ${\bf 1}$. 
Since the trace is symmetric, the commutator of two elements in \hs{\lambda} then does not involve
${\bf 1}$, and hence the Lie algebra is a direct sum of $\hs{\lambda}\oplus {\mathbb C}$. 

To get a feeling for the structure of \hs{\lambda} it is useful to work out the first few terms explicitly. 
For example we have 
\be\label{V3}
\begin{array}{ll}
V^3_2  = J_+ J_+ \ , \qquad V^3_{-2} = J_{-} J_{-} \qquad  & 
V^3_0 = \frac{1}{3} \bigl( J_{-} J_{+} + J_0 + 2 J_0 J_0 \bigr) 
\cong J_0 J_0 - \frac{1}{12}(\lambda^2-1)\\
V^3_1 = J_0 J_+ + \frac{1}{2} J_+\ , \qquad 
& V^3_{-1} =  J_{-} J_0 + \frac{1}{2} J_{-} \ ,
\end{array}
\ee
and the first few commutators are 
\be
\begin{array}{ll}
{}[V^3_2,V^3_1] = 2 \, V^4_3 \qquad 
&[V^3_2,V^3_0] = 4\, V^4_2 \\
{}[V^3_2,V^3_{-1}] = 6\, V^4_1 - \tfrac{1}{5} (\lambda^2-4) \, V^2_1 \qquad
&[V^3_2,V^3_{-2}] = 8\, V^4_0 - \tfrac{4}{5} (\lambda^2-4)\, V^2_0 \ .
\end{array}
\ee
It is easy to check that these coefficients agree with (\ref{vcom}). 
These identities suggest that for $\lambda=2$, the Lie algebra generated by 
$V^s_n$ with $s\geq 3$ form a proper subalgebra of \hs{\lambda}. In fact, this is a special
case of a more general phenomenon. If $\lambda=N$ with integer $N\geq 2$
then the quadratic form (\ref{quadform}) degenerates 
\cite{Feigin88,Vasiliev:1989re,Fradkin:1990qk},
\be
\tr(V^s_m V^r_n) = 0  \quad \mbox{for}  \quad s>N \ .
\ee
This implies that an ideal $\chi_N$ appears, consisting of all generators $V^s_n$ 
with $s>N$.  Factoring over this ideal truncates to the finite algebra sl$(N)$,
\be\label{slN}
\mbox{sl}(N) = \hs{N}/\chi_N  \quad (N\geq 2) \ .
\ee
Note that when $\lambda=1$, the quadratic form (\ref{quadform})
vanishes identically (as is obvious from its definition). However, one can
rescale the trace by $1/(\lambda-1)$, and the result is non-degenerate.
In fact, the resulting Lie algebra then agrees with the wedge subalgebra of $\W_{\infty}^{\rm PRS}$
as defined by Pope, Romans and Shen \cite{Pope:1989ew}, see also \cite{Pope:1990rn} and 
references therein. (As will be explained in more detail below, the $\W$-algebra 
$\W_{\infty}^{\rm PRS}$ is linear, and hence the modes $W^s_n$ with $|n|< s$ form a 
subalgebra, which agrees with $\hs{1}$.) Note that $\lambda$ is the 't~Hooft parameter of \cite{Gaberdiel:2010pz}, so this case corresponds to the maximal coupling limit.

We should also mention that for $\lambda=\frac{1}{2}$, the algebra is
isomorphic to the hs$(1,1)$ algebra as defined in \cite{Fradkin:1986ka,Blencowe:1988gj,Bergshoeff:1989ns}. 
Blencowe \cite{Blencowe:1988gj} defined the original theory of higher-spin AdS$_3$ gravity 
as a Chern-Simons theory with bosonic subalgebra hs$(1,1)\;\oplus\;$hs$(1,1)$.  This is 
also the bulk algebra considered recently by Henneaux and Rey \cite{Henneaux:2010xg}. 

In the limit $\lambda \to \infty$ the commutator 
 algebra of sl$(2)$ used in the construction (\ref{idealcon}) reduces to a classical Poisson 
 bracket algebra \cite{Bergshoeff:1989ns,Pope:1989cr}.  The sl$(2)$ generators $J_0, J_{\pm}$ (properly rescaled) can be considered 
 coordinates on the 2d hyperboloid defined by $C_2=1$.  The \hs{\infty} algebra is then the 
 Lie-bracket algebra of area-preserving diffeomorphisms of the hyperboloid $H^2$.  
 This can also be defined as an $N\to \infty$ limit of sl$(N)$, so
\be
\hs{\infty} = s\mbox{diff}(H^2) = \mbox{sl}(\infty) \ .
\ee

Finally, we should mention that for any value of $\lambda$ the zero modes $V^s_0$ all commute
with one another. (This follows because in the quotient space $B[\mu]$ we may represent them
in terms of polynomials of $J_0$.) Thus these algebras contain infinitely many commuting charges. 

\subsection{$\W$-algebras and Wedge Algebras}\label{ss:wedge}

$\W$-algebras consist of the Virasoro generators $L_n\equiv W^2_n$ at level 2, 
plus a tower of higher-dimension currents. For example, the family $\W_N$ has primary 
operators of dimension $s$ for $s=2,3,\dots,N$.  Its commutation relations are nonlinear, 
so it is not a Lie algebra (when expressed in terms of these modes). The most familiar example
is the famous  $\W_3$ algebra of  \cite{Zamolodchikov:1985wn}, for which the commutation relations 
are 
\bea\label{w3}
\ [W^2_m, W^2_n] &= &  \tfrac{c}{12}\, m \, (m^2-1)\, \delta_{m,-n} + (m-n)\,  W^2_{m+n} \\
\ [W^2_m, W^3_n] & = & (2m-n)\, W^3_{m+n}\notag\\
\ [W^3_m, W^3_n] &= &  \tfrac{c}{3 \cdot 5!}\, (m^2-2)\, (m^2-1)\, m \, \delta_{m,-n}\ + 
\tfrac{1}{30} \, (m-n)  (2m^2-mn+2n^2-8) W^2_{m+n}  \notag \\
& & 
+ \tfrac{16}{22+5c} (m -n) \Bigl(\sum_{p\in \mathbf{Z}} :W^2_{m+n+p}W^2_{-p}: + \, x_{m+n}\, W^2_{m+n} \Bigr) 
\ , \notag
\eea
where the $x_{m}$ are some constants. The commutation relations are only known explicitly for 
$\W_N$ with $N=3$ and $N=4$. 

Because of the first term in the second line of $[W^3_m, W^3_n]$, the $\W$-algebra is not a (linear)
Lie algebra. As a consequence the definition of the `finite' or `global' subalgebra requires some care. 
Naively, this subalgebra should consist of the modes that annihilate the vacuum, {\it i.e.}\ it should be
generated by the `wedge' modes $W^s_n$ with $|n|<s$. However, because of the non-linear term, 
the commutator of $W^3_2$ with $W^3_1$, say, contains terms involving $W^2_p$ with $p$ arbitrary. 
Thus the above brackets do not close on the wedge modes. 

Given the structure of the above algebra, it is not difficult to see how this can be repaired, at least
in this case: the non-linear terms decouple if we take $c\rightarrow \infty$, and the central terms
do not contribute provided that we restrict ourselves to the wedge modes. For the case at hand, the
wedge algebra is then simply
\bea\label{w3wedge}
\ [V^2_m, V^2_n] &= &  (m-n)\,  V^2_{m+n} \\
\ [V^2_m, V^3_n] & = & (2m-n)\, V^3_{m+n}\notag\\
\ [V^3_m, V^3_n] &= & \tfrac{1}{30} \, (m-n)  (2m^2-mn+2n^2-8) V^2_{m+n} \ , \notag
\eea
for $|n|<s$, which is easily seen to be isomorphic to sl$(3)$. 

The above construction was generalised by Bowcock and Watts \cite{Bowcock:1991zk} 
under some fairly mild conditions to general $\W$-algebras. In particular, they showed that the wedge 
algebra of  $\W_N$ is sl$(N)$. More generally, if $\W(\mathfrak{g})$ is the $\W$-algebra constructed from a 
Lie algebra $\mathfrak{g}$ by Drinfeld-Sokolov reduction \cite{Drinfeld:1984qv,Balog:1990dq,Bouwknegt:1992wg} of the affine algebra 
$\hat{\mathfrak{g}}$, then the procedure of Bowcock and Watts gives back $\mathfrak{g}$,
\be
\mathfrak{g} \xrightarrow{\mbox{\small Drinfeld-Sokolov}} 
\W(\mathfrak{g}) \xrightarrow{\mbox{\small Bowcock-Watts}} \mathfrak{g} \ .
\ee
The analysis of Bowcock and Watts was concerned with finite-dimensional Lie algebras 
$\mathfrak{g}$.  In the current context, we are interested in the analogous statement for
$\mathfrak{g}=\hs{\lambda}$. While a direct application of the Drinfeld-Sokolov reduction in this case
is somewhat delicate, one can think of the Drinfeld-Sokolov reduction as describing the asymptotic
symmetries of the corresponding Chern-Simons gravity theory. Thus we can {\em determine}
\w{\lambda} starting from \hs{\lambda} by analyzing the algebra of asymptotic symmetries. 
In this context, the non-linearities will be related to the curvature of
AdS$_3$, and hence should disappear in the limit where the cosmological constant goes to zero,
{\it i.e.}\ for $c\rightarrow \infty$. Thus we expect that we can reobtain from \w{\lambda} the
original higher spin algebra \hs{\lambda} by going to the wedge, {\it i.e.}\ that we have 
\be
\hs{\lambda} \xrightarrow{\mbox{\small Asymptotic symmetries}} \w{\lambda} \xrightarrow{c\to \infty \ , \  |n|<s} \hs{\lambda} \ .
\ee
This is confirmed in Section \ref{s:asg}.  By construction, \w{\lambda} describes the symmetries of the boundary CFT, while \hs{\lambda} is associated to the 
bulk symmetries. The physical origin behind this algebraic statement is therefore the usual relation between
bulk and boundary symmetries. 

In the following we want to construct \w{\lambda} explicitly by calculating the
Poisson brackets of the corresponding charges. 
Because it is realized by Poisson brackets, it is a classical $\W$-algebra as 
in \cite{Henneaux:2010xg,Campoleoni:2010zq}; for example, for $\W_3$ it 
differs from (\ref{w3}) by the quantum correction that shifts the 
denominator of the last term of (\ref{w3}), $5c \to 5c + 22$. Note that this does not affect the 
wedge algebra. In Section \ref{ss:CFT}, we shall show that the 
primary fields of the dual CFT proposed in \cite{Gaberdiel:2010pz} indeed 
define representations of \w{\lambda}; this analysis essentially only relies on the structure
of the wedge algebra \hs{\lambda}.

\section{\w{\lambda} from Asymptotic Symmetries }\label{s:asg}

In this section we generalize the asymptotic symmetry analysis of 
\cite{Henneaux:2010xg,Campoleoni:2010zq} to a bulk theory of higher-spin gravity based 
on the algebra \hs{\lambda}.  The results for sl$(N)$ \cite{Campoleoni:2010zq} and hs$(1,1)$ 
\cite{Henneaux:2010xg} can be recovered by setting $\lambda$ to special values as described 
in Section~\ref{ss:hsalgebra}.

The asymptotic symmetry algebra is the algebra of allowed, nontrivial symmetries of the theory.  
A symmetry is `allowed' if it generates a transformation obeying the boundary conditions; it is 
`nontrivial' if the associated conserved charge is nonzero.  Because conserved charges are given 
by an integral over the boundary of a spatial slice, nontrivial symmetries are those that act at infinity.  
The commutation relations of the algebra follow from the Poisson bracket algebra of conserved charges. 

Before doing the detailed computation, let us describe the general structure that we expect.  In any 
theory, the algebra of conserved charges is identical to that of the symmetries themselves, up to a 
possible extra term,
\be\label{generalq}
 \{ Q(\Lambda), Q(\Gamma) \} = Q([\Lambda,\Gamma]) + K(\Lambda, \Gamma) \ ,
\ee
where $\Lambda,\Gamma$ are gauge parameters, $Q$ is the conserved charge, and $K$ 
denotes the extra term.  If the allowed gauge parameters 
$\Lambda,\Gamma$ are field-independent, then $K$ is a $Q$-independent central term.  For pure 
gravity in AdS$_3$, the allowed diffeomorphisms are indeed independent of the metric and $K$ leads to the Brown-Henneaux central charge of the boundary Virasoro algebra \cite{Brown:1986nw}.  More generally, the allowed gauge transformations can be field dependent, in which case $K$ may depend
nonlinearly on $Q$,
\be
K(\Lambda,\Gamma) \sim \mbox{nonlinear terms in $Q$} + \mbox{central terms} \ .
\ee
For global symmetries of the bulk vacuum, in our case \hs{\lambda}, the term 
$Q([\Lambda, \Gamma])$ in (\ref{generalq}) is again a generator of \hs{\lambda}, while 
$K$ has only nonlinear contributions.  Thus the asymptotic symmetries form a nonlinear algebra 
whose linear, global part is \hs{\lambda}, directly paralleling the discussion of 
section~\ref{ss:wedge}. (The gauge-fixing procedure used to simplify the computation below 
obscures this relationship by introducing additional field dependence in the gauge transformations, 
but the structure (\ref{generalq}) is guaranteed to reappear in the final answer.)

To compute the asymptotic algebra we apply the formalism of \cite{Regge:1974zd,Brown:1986nw, Coussaert:1995zp,Banados:1998pi,Henneaux:1999ib} to the 
Chern-Simons formulation of higher spin gravity.  This is a topological theory with only boundary 
excitations.  Although for holographic minimal models we are ultimately interested in a theory containing 
additional propagating scalars, we do not expect the scalars to affect the asymptotic symmetries for 
masses above the Breitenlohner-Freedman bound.  The case $M^2 = M_{\rm BF}^2$, or $\lambda = 0$ 
in the dual CFT, may allow for interesting modifications because the scalars have relaxed behavior at 
infinity \cite{Henneaux:2002wm,Henneaux:2004zi}.  Here we simply impose boundary conditions on 
the scalars that prevent any new contribution to the asymptotic charges.

\subsection{The Bulk Theory}
We consider a theory of higher spin gravity in AdS$_3$ given by the Chern-Simons action
\be
S = S_{\rm CS}[A] - S_{\rm CS}[\tilde{A}]\ ,
\ee
with
\be
S_{\rm CS}[A] = {\hat{k}\over 4\pi}\int \tr\left( A\wedge dA + {2\over 3}A\wedge A \wedge A\right) \ .
\ee

The level $\hat{k}$ here is related, but not equal, to the level $k$ of the coset CFT on the boundary,
see \cite{Gaberdiel:2010pz}.
$(A,\tilde{A})$ take values in\footnote{The bulk analysis is well defined for all $\lambda\geq 0$, but to 
compare safely to the CFT one should restrict to $0 < \lambda < 1$.}
\be
\hs{\lambda}\oplus \hs{\lambda} \ ,
\ee
with generators
\be
V^s_n \ , \quad |n|<s \ ,\qquad \tilde{V}^t_m \ ,  \quad |m| <t \ .
\ee
This theory was first defined in \cite{Bergshoeff:1989ns}, and the trace was 
derived in \cite{Vasiliev:1989re}, though we use the PRS formulation of the higher spin 
algebra \cite{Pope:1989sr} as described in Section \ref{ss:hsalgebra}.  
We focus on $A$, but similar statements hold for $\tilde{A}$.

Following \cite{Henneaux:2010xg,Campoleoni:2010zq}, imposing the AdS 
boundary conditions  and gauge fixing sets 
\be
A_+ = e^{-\rho V^2_0} a(T+\phi) e^{\rho V^2_0} \ , \quad A_- = 0 \ , \quad 
A_\rho = e^{-\rho V^2_0} \p_\rho e^{\rho V^2_0} \ ,
\ee
where $T,\phi$ are boundary coordinates, $\rho$ is the radial coordinate, $A_\pm = A_\phi \pm A_T$, and
\be\label{hwgauge}
a(T+\phi) = V^2_1 + {2\pi \over \hat{k}} \sum_{s\geq 2} {1\over  N_s} L_s(T+\phi) V^s_{-s+1}  \  .
\ee
The $L_s$ are arbitrary functions which will be interpreted as currents of the dual CFT.  From 
now on we work at fixed time $T$, so $a = a(\phi)$, $L_s = L_s(\phi)$. The coefficient $N_s$ chosen to 
normalize the currents in (\ref{hwgauge}) is
\be\label{defN}
N_s = \tr(V^s_{-s+1} V^s_{s-1})\ .
\ee
$N_s$ is a rational function of $\lambda$, see (\ref{appN}) for the explicit formula.

\subsection{Gauge Transformations}
The \hs{\lambda} gauge symmetries of the form
\be
\Gamma(T+\phi) = e^{-\rho V^2_0} \gamma(T+\phi) e^{\rho V^2_0} 
\ee
preserve the gauge fixing condition, and act on the gauge field as
\be\label{deltaa}
\delta a = \gamma' + [a,\gamma] \ .
\ee
At fixed time $T=0$, let us expand them in components as 
\be\label{lambdacomp}
\gamma(\phi) = \sum_{s\geq 2}\sum_{|n|<s} \gamma_{s,n}(\phi) V^s_n \ .
\ee
The highest-weight generators play a special role so we denote them by
\be
\eta_s(\phi) \equiv  \gamma_{s,s-1}(\phi)   \ .
\ee

Gauge symmetries which do not vanish near the boundary are physical symmetries, 
relating physically inequivalent states.  These symmetries have corresponding conserved 
charges\footnote{For field-independent gauge transformations, this would be 
$Q(\gamma) = {\hat{k}\over 2\pi}\int d\phi\, \tr(a\gamma)$.  When the gauge parameter is 
field-dependent, the $V^2_1$ component of $a$ must be dropped to ensure that the variation 
of this charge cancels the boundary term in the variation of the bulk generator of field-dependent 
gauge transformations.}
\be\label{finalq}
Q(\gamma) = \int d\phi \sum_{s\geq 2} L_s\eta_s \ .
\ee
These charges generate gauge transformations under Poisson brackets.  That is, for any expression 
$X$ given in terms of the phase space variables,
\be\label{poissonvar}
\{ X  \ , Q(\gamma)   \} = \delta_\gamma X \ .
\ee

The gauge parameters $\gamma_{s,n}$ are not all independent, because we must also 
restrict to gauge transformations that maintain the boundary condition (\ref{hwgauge}).  
Plugging a general $\gamma$ (\ref{lambdacomp}) into the transformation law (\ref{deltaa}),
and using the \hs{\lambda} commutation relations (\ref{vcom}) gives
\be\label{deltaac}
\delta a = \sum_{r \geq 2}\sum_{|n|<r} c_{r,n} V^r_n
\ee
with
\be\label{coV}
c_{r,n} = \gamma_{r,n}' + (-n+r)\,\gamma_{r,n-1} 
+  \sum_{s \geq 2}\sum_{u} {2\pi\over \hat{k}  N_s}L_s\, \gamma_{r+u-s,n+s-1}\, 
g^{s,r+u-s}_{u}(-s+1,n+s-1;\lambda) \ . 
\ee
The term containing $\gamma_{r,n-1}$ should be dropped if $n=-r+1$, and the range of $u$ is
\be\label{rangeu}
2s > u \geq \max(2, s-r+|n+s-1|+1)  \ , \quad u\in 2\mathbf{Z} \ .
\ee
The boundary condition (\ref{hwgauge}) requires
\be\label{crn}
c_{r,n} = 0 \ , \quad n \neq -r+1 \ .
\ee
This infinite set of equations for the gauge parameters $\gamma_{r,n}$ can be solved iteratively in terms of the highest-weight gauge parameters $\eta_r$ \cite{Henneaux:2010xg}.  

The existence of the conserved charges (\ref{finalq}) indicates that the asymptotic symmetries form a 
$\W$-algebra, with one current at each spin $s \geq 2$.  The commutation relations of the $\W$-algebra,
\be
\{L_s(\phi), L_t(\phi') \} \ ,
\ee
for given choices of $s,t$ are computed as follows.  First, keep $\eta_s$ arbitrary 
but set all other $\eta_r = 0$, $r\neq s$. Solve the equations (\ref{crn}) iteratively, 
fixing the $\gamma_{r,n}$ as functions of the gauge parameter $\eta_s$. Then from 
(\ref{hwgauge}) and (\ref{deltaac}), the asymptotic symmetry algebra is
\be\label{deltac}
\delta_s L_t = {\hat{k}\over 2\pi}N_t c_{t,-t+1}(\eta_s) \ .
\ee
This variation can be converted to a Poisson bracket using (\ref{poissonvar}) and (\ref{finalq}),
\be\label{getcom}
\int d\phi\ \eta_s(\phi) \{ L_t(\phi'), L_s(\phi) \} = \delta_s L_t(\phi') 
= {\hat{k}\over 2\pi} N_t  c_{t,-t+1}(\eta_s(\phi'))  \ .
\ee

\subsection{The Structure of \w{\lambda}}

We now apply the procedure described above to compute structure constants of \w{\lambda}. 
We write the results for $\delta_s L_t$ to keep the formulae compact, but this can easily be 
converted to the commutator of currents using (\ref{getcom}), or into commutators for the Fourier modes.  
For now we give only the results for low spins, but arbitrary spins will be considered from another angle 
in section \ref{ss:fmr}.

As an example, first consider the action of the stress tensor $s=2$.  All $\lambda$-dependence drops out, 
so this was computed in \cite{Henneaux:2010xg}. The solution of (\ref{crn}) for $r>2$ is
\bea
\gamma_{r,n} &=& 0 \quad (n>-r+1) \\
\gamma_{r,-r+1} &=& {2\pi\over \hat{k}  N_r}L_r \eta_2 \ , \notag
\eea
while for $r=2$,
\be
\gamma_{2,0} = -\eta_2' \ , \quad \gamma_{2,-1} = {2\pi\over \hat{k} N_2}L_2 \eta_2 + \half \eta_2'' \ .
\ee
Plugging these into $c_{r,-r+1}$ and using (\ref{deltac}) gives the $\delta_2 L_r$ variations written explicitly below.  

More generally, the solutions for $\gamma_{t,n}$ depend on the structure constants $g^{st}_u(m,n;\lambda)$ and therefore involve polynomials in $\lambda$.  The results for low spins are
\bea
\delta_2 L_2 &=& L_2' \eta + 2 \eta' L_2 - {c\over 24 \pi} \eta''' \label{com22}\\
\delta_2 L_t &=& L_t'\eta + t \eta' L_t  \qquad \qquad (t>2)\label{com2t}\\
\delta_3 L_3 &=&  4 L_4 \eta' + 2L_4'\eta-{N_3\over 12}(15L_2'\eta''+9\eta'L_2'' + 10 L_2 \eta''' + 2 L_2'''\eta)  \label{com33} \\
& & \quad + {32\pi\over c}N_3 L_2(L_2\eta)' + {c\over 288 \pi}N_3 \p_\phi^5\eta\notag\\
\delta_3 L_4 &=& 5 L_5 \eta' + 2 \eta L_5'+ {N_4\over 15 N_3}(14L_3'\eta'' + 6 L_3''\eta' + 14 L_3 \eta''' + \eta L_3''')\\ 
& & \quad - {8\pi\over 5 c} {N_4\over N_3}(25L_3L_2' \eta+ 52 L_2 L_3 \eta' + 18 L_2 L_3' \eta)\notag\\
\delta_3 L_5 &=& 6 L_6 \eta'+ 2 \eta L_6' + {N_5\over 56 N_4}(45 L_4'\eta'' 
+ 15 L_4'' \eta' + 60 L_4 \eta''' + 2 L_4''' \eta) \\
& & \quad + {84\pi \over 5 c}{N_5\over N_3^2}(3L_3^2 \eta' + 2 L_3 L_3' \eta) 
- {48\pi\over 7 c}{N_5\over N_4}(7L_4 L_2' \eta + 15 L_4 L_2 \eta' + 4 L_2 L_4' \eta)\ .\notag
\eea
The schematic form of the first spin-4 variation is
\bea
\delta_4 L_4 &\sim& L_6 \eta + N_4 L_2 \eta + g^{44}_4(-3,0;\lambda) L_4 \eta 
+ {N_4\over c} (L_2)^2 \eta + {560N_4^2+69N_3 N_5 \over c N_3^2} (L_3)^2\eta \notag\\
& & \quad + \  {1\over c}{g^{44}_4(-3,0;\lambda)}L_4 L_2 \eta + {N_4 \over c^2}(L_2)^3 \eta 
+ {c N_4\over 8640\pi} \p_\phi^7\label{com44} \eta \ ,
\eea
where we have ignored derivatives and numerical factors except in the central term, 
but kept all $\lambda$ and $c$ dependence.  The subscript on $\eta$ has been suppressed; 
for spin-$s$ variations $\delta_s L_t$, take $\eta$ to be $\eta_s$. 
Reversed variations, like $\delta_3 L_2$, can be found by the same process, 
or by converting (\ref{com22})--(\ref{com44}) to Poisson brackets using (\ref{getcom}),
and then using (\ref{getcom}) again to compute the variation; for example, 
\be
\delta_3 L_2  = 3 L_3 \eta' + 2 L_3' \eta \ .
\ee
The first few $\lambda$-dependent factors, defined in (\ref{defN}) and (\ref{specialg}), are
\bea
N_3 &=& {16\over 5}q^2(\lambda^2-4)\\
N_4 &=& -{384\over 35}q^4(\lambda^2-4)(\lambda^2-9) \notag \\
N_5 &=& {4096\over 105}q^6(\lambda^2-4)(\lambda^2-9)(\lambda^2-16) \notag \\
g^{44}_4(-3,0;\lambda)  &=&{16\over 5}q^2(\lambda^2-19) \notag \ .
\eea
(Recall $q$ is a normalization factor in the algebra that can be set to one.)  

The central charge of the Virasoro algebra $\delta_2 L_2$ is
\be
c = 6 \hat{k} \ .
\ee
According to (\ref{com2t}), the higher spin currents are Virasoro primaries.  

It is straightforward to convert to modes.  To put the algebra in standard form with no central 
terms inside the wedge $|n| < s$, we first shift the stress tensor
\be
L_2(\phi) \to L_2(\phi) - {\hat{k}\over 8\pi} \ .  
\ee
We have checked that if 
we restrict to the wedge $|n| <s$ and scale $c \to\infty$ to eliminate the nonlinear terms, 
then the variations above become the commutators (\ref{vcom}) of \hs{\lambda} as required by 
the general discussion around (\ref{generalq}).  This is the Bowcock-Watts procedure applied 
to the infinite Lie algebra \hs{\lambda}, and demonstrates explicitly the relationship between
 \w{\lambda} and \hs{\lambda}.

\subsection{Linear $\W_\infty[\lambda]$ at $\lambda=1$}\label{ss:linear}
The \w{\lambda} algebra given above appears to be nonlinear, since quadratic and higher 
terms appear on the right-hand side. This nonlinearity stems from the fact that the allowed 
gauge transformations are field dependent.  By contrast in ordinary AdS$_3$ gravity the 
Brown-Henneaux diffeomorphisms are fixed once and for all, and do not depend on the metric.

To confirm that \w{\lambda} is truly nonlinear, we must check whether a redefinition of the currents can 
linearize the algebra.  For example, in the $[L_3, L_3]$ commutation relation (\ref{com33}), we can 
absorb the nonlinear term $(L_2)^2$ into a redefinition of $L_4$.  There is no guarantee, however, 
that this will work for higher commutators.  We will show that for generic $\lambda$, \w{\lambda} is indeed 
truly nonlinear, but that for $\lambda = 1$  the algebra linearizes after redefining the generators.  In 
fact, it becomes the linear algebra $\W_\infty^{\rm PRS}$ defined by Pope, Romans and 
Shen \cite{Pope:1989ew} (see also \cite{Pope:1990rn} and references therein)
\be
\W_\infty[1] = \W_{\infty}^{\rm PRS} \ .
\ee
The wedge algebra of $\W_\infty^{\rm PRS}$ (which in this case is actually a proper 
subalgebra since there are no nonlinearities) is \hs{1}.  In fact, the full commutation relations of 
$\W_\infty^{\rm PRS}$ are given by (\ref{vcom}) with $\lambda=1$, where we now 
allow $m,n$ to range over all integers instead of restricting to the wedge modes.  
For $\lambda\neq 1$, the \hs{\lambda} commutation relations (\ref{vcom}) cannot be extended 
outside the wedge in this manner, as the resulting algebra would violate the Jacobi 
identity.\footnote{If the spin-1 current is included, then \hs{0}$\oplus \mathbb{C}$ can also be 
extended outside the wedge, resulting in the $\W_{1+\infty}^{\rm PRS}$ algebra constructed in  
\cite{Pope:1990kc}.  $\W_{1+\infty}^{\rm PRS}$ is related to \w{0} by a constraint that removes 
the spin-1 current, introducing nonlinearities through the Dirac bracket procedure 
\cite{FigueroaO'Farrill:1992cv}; this is a special case of the general construction of \w{\lambda} 
described in section \ref{ss:fmr} below.  For other values of $\lambda$, \hs{\lambda} can be 
embedded in a linear $\W_\infty$ algebra by twisting $\W_{1+\infty}^{\rm PRS}$ 
\cite{Pope:1990be,Bergshoeff:1990cz,Bergshoeff:1991dz}, but the resulting algebras have no 
obvious connection to the nonlinear algebra constructed here.} Thus it is not surprising that we 
find a linear algebra at $\lambda=1$ and a nonlinear algebra otherwise.

The linearization of the algebra relies on a large number of nontrivial cancellations in the commutators, 
which we have checked for $\{L_3, L_3\}$, $\{L_4, L_3\}$, $\{L_5, L_3\}$, $\{L_4, L_4\}$ and $\{L_5, L_4\}$, 
fixing the redefined generators through spin 7. These commutators already greatly overconstrain the 
field redefinitions required to define a linear algebra, so this strongly suggests that linearity continues 
to all orders.

The redefined generators will be denoted $\tilde{L}_s$.  No redefinitions are necessary for 
spin-$2,3$ other than a shift of the zero mode:
\be
L_2 = \tilde{L}_2 - {\hat{k}\over 8 \pi}\  , \quad L_3 = \tilde{L}_3 \ .
\ee
Consider $\{\tilde{L}_3, \tilde{L}_3\}$ given in (\ref{com33}). This is linear if we redefine
\be\label{redef4}
L_4 = \tilde{L}_4 + \beta \tilde{L}_2^2 \ , \qquad 
\beta = -{64 \pi\over  15\hat{k}}q^2(\lambda^2-4) \ .
\ee
Now consider 
\be
 \{\tilde{L}_4, \tilde{L}_3\} = \{L_4 , L_3 \} - 2 \beta \tilde{L}_2 \{L_2 \ , L_3 \} \ .
\ee
The nonlinear terms must be absorbed into a redefinition of $L_5$ of the form
\bea
L_5 &=& \tilde{L}_5 + \gamma \tilde{L}_2 \tilde{L}_3 \ .
\eea
Plugging into (\ref{com33}) and (\ref{com2t}) we find 
\be
\beta = -{8 \pi \over 5 \hat{k}}q^2(\lambda^2-9)\ , \quad \gamma = \frac{50}{7} \beta \ .
\ee
Comparing the restriction on $\beta$ to (\ref{redef4}), we see that the algebra cannot be 
linearized unless $\lambda=1$.  Thus we set $\lambda=1$ and proceed, with coefficients so far
\be\label{betagamma}
\beta = {64\pi q^2\over 5\hat{k}} \ , \quad \gamma = {640 \pi q^2\over 7 \hat{k}} \ .
\ee
Next consider $\{\tilde{L}_4 \ , \tilde{L}_4\}$.  Using (\ref{getcom}), this is related to the variation 
\be\label{var44}
\tilde{\delta}_4 \tilde{L}_4 = \delta_4 L_4 + \delta_2 L_4 
- 2 \beta \tilde{L}_2(\delta_4 L_2 + \delta_2 L_2)
\ee
where we set
\be
\eta_4 = \eta \ , \quad \eta_2 = -2\beta \eta \tilde{L}_2 \ .
\ee
The right-hand side of (\ref{var44}) has 10 nonlinear terms, but all vanish at $\lambda=1$ if we choose
\bea\label{redef6}
L_6 &=& \tilde{L}_6 + {40960 \pi^2 q^4\over 21 \hat{k}^2} \tilde{L}_2^3 + {640\pi q^2\over 3\hat{k}} \tilde{L}_2\tilde{L}_4 
+ {5440\pi q^2\over 21\hat{k}} \tilde{L}_3^2 - {1280 \pi q^4\over 21\hat{k}} (\tilde{L}_2')^2 \\
& & \quad + \ {1024 \pi q^4\over 21 \hat{k}} \tilde{L}_2\tilde{L}_2'' - {1024 \pi q^4\over 21\hat{k}}\tilde{L}_2^2 \ . \notag
\eea
Similarly, the variation corresponding to $\{\tilde{L}_5, \tilde{L}_4 \}$ has 17 nonlinear terms which disappear at $\lambda=1$ if we define
\bea\label{redef7}
L_7 &=& \tilde{L}_7 + {4480\pi q^2\over 11\hat{k}} \tilde{L}_5 \tilde{L}_2 
+ {58240\pi q^2\over 33\hat{k}} \tilde{L}_4 \tilde{L}_3 - {17920\pi q^4\over 33\hat{k}} \tilde{L}_2' \tilde{L}_3' \\
& & \quad + {3584\pi q^4\over 11\hat{k} }\tilde{L}_2'' \tilde{L}_3
+ {5120\pi q^4\over 33 \hat{k}} \tilde{L}_2 \tilde{L}_3'' + {1433600\pi^2 q^4\over 33\hat{k}^2} \tilde{L}_2^2 \tilde{L}_3 
-{6144 \pi q^4\over 11\hat{k}}\tilde{L}_2 \tilde{L}_3 \ .\notag
\eea
Having fixing the $\tilde{L}_{s}$ generators 
for $s\leq 7$, we have fully determined all currents appearing in $\{\tilde{L}_5 , \tilde{L}_3\}$.  
This is also linear. Note that inhomogeneous terms, like the last term in (\ref{redef6}) and 
(\ref{redef7}), are allowed when going to a nonprimary basis and result in the expected 
$\{\tilde{L}_2, \tilde{L}_6\}$ and $\{\tilde{L}_2, \tilde{L}_7\}$ commutators.

Converting the $\tilde{\delta}_s \tilde{L}_t$ variations to mode commutators 
gives precisely the $\W_\infty^{\rm PRS}$ algebra discussed above.  This is the natural 
linear extension of \hs{1} outside the wedge, and provides a consistency check of our computation 
because $\W_\infty^{\rm PRS}$ is known to satisfy the Jacobi identity.

\subsection{Full Commutation Relations at arbitrary $\lambda$}\label{ss:fmr}
So far we have resorted to case-by-case computations at low spins, rather than attempting 
to find a general solution to the infinite system of equations (\ref{coV}) determining the asymptotic algebra.  
We will now demonstrate that after a change of basis, the low-spin commutation relations exactly match a 
one-parameter family of nonlinear $\W_\infty$ algebras discovered by 
Figueroa-O'Farrill, Mas, and Ramos \cite{FigueroaO'Farrill:1992cv} and by 
Khesin and Zakharevich \cite{Khesin:1993ru,Khesin:1993ww}.
The full commutation relations of this algebra 
are known, so this provides the explicit commutators of 
\w{\lambda} for all spins in closed form.

In \cite{FigueroaO'Farrill:1992cv,Khesin:1993ru,Khesin:1993ww}, building on 
\cite{Bakas:1991fs,Bakas:1991gs,Yamagishi:1991ax,FigueroaO'Farrill:1991ek,Yu:1991ng,Yu:1991bk}, 
a family of non-linear $\W_\infty$ algebras was proposed in the context of integrable systems and the 
KP hierarchy (a generalization of the KdV hierarchy).  The construction starts with a one-parameter nonlinear 
algebra $\W_{\rm KP}^{(\lambda)}$ with currents $U_s$ of dimensions $s=1,2,\dots$. 
(The algebra $\W_{\rm KP}^{(\lambda)}$ is a Hamiltonian structure for the KP hierarchy and can be realized by pseudodifferential operators.)   
Imposing the second-class 
constraint $U_1=0$ and going to the induced Dirac brackets gives a nonlinear $\W$-algebra of spins
$2,3,\dots$, denoted $\hat{\W}_\infty^{(\lambda)}$ in \cite{FigueroaO'Farrill:1992cv}.  We claim that this 
algebra is identical to the  asymptotic symmetry algebra of higher spin gravity,
\be\label{fmrid}
\hat{\W}_\infty^{(\lambda)} \cong \w{\lambda} \ .
\ee
The first evidence for this isomorphism comes from the degeneration points.  It was observed in 
\cite{FigueroaO'Farrill:1992cv,Khesin:1993ru,Khesin:1993ww} that $\hat{\W}_\infty^{(\lambda)}$ is a 
`universal' $\W$-algebra, in the sense that other known $\W$-algebras can be obtained by setting $\lambda$ 
to specific values.  Setting $\lambda=1$ gives $\W_\infty^{\rm PRS}$, while setting $\lambda=N$ for 
integer $N\geq 2$ and constraining fields with spins greater than $N$ to vanish leads to $\W_N$.  
The same is true for \w{\lambda}; the case $\lambda=1$ was shown in section \ref{ss:linear} and the 
case $\lambda=N$ follows from (\ref{slN}) together with the results of \cite{Campoleoni:2010zq}.

This is suggestive, but to identify the two algebras we must compare the commutation 
relations as a function of $\lambda$.  The process is similar to checking linearity in section 
\ref{ss:linear} so we will be brief.  The $\hat{\W}_\infty^{(\lambda)}$ commutators are 
\cite{FigueroaO'Farrill:1992cv}
\be\label{fmrcom}
\ \{U_s(\phi) \ , \ U_t(\phi') \} = P_{st}(\phi)\delta(\phi - \phi') \ ,
\ee
where $P_{st}$ is a differential operator given in appendix \ref{a:fmr}. This can be converted to 
variations $\delta_s U_t$ using the first equality in (\ref{getcom}).  The action of the stress tensor 
$U_2$ on higher spin fields indicates that they are not Virasoro primary, whereas we computed 
\w{\lambda} in a primary basis.  Therefore to compare with the \w{\lambda} algebra 
(\ref{com22})--(\ref{com44}), we first go to a non-primary basis,
\be
L_2 = U_2 \ , \quad L_3 = U_3 + p_3 U_2' \ , \quad L_4 = U_4 + p_4 U_2^2 + p_5 U_2'' + p_6 U_3' \ , 
\quad \dots.
\ee
Plugging this ansatz into (\ref{com22})--(\ref{com44})
and choosing the coefficients to reproduce the $\{U_2  , U_s\}$ commutators in (\ref{fmrcom}) 
fixes all of the coefficients $p_i$ for spins $s \leq 5$, 
\bea 
L_2 &=& U_2\\
L_3 &=& U_3 - \half(\lambda-2)U_2' \notag\\
L_4 &=& U_4 + \half(3-\lambda)U_3' 
+ {1\over 10}(\lambda-2)(\lambda-3)U_2'' 
- {(\lambda-2)(\lambda-3)(5\lambda+7)\over 10 c (\lambda^2-1)}U_2^2 \notag\\
L_5 &=& U_5  -\half(\lambda-4)U_4' 
+ {3\over 28}(\lambda-3)(\lambda-4)U_3''
-{1\over 84}(\lambda-2)(\lambda-3)(\lambda-4)U_2''' \notag\\
& & \quad + \ {(\lambda-3)(\lambda-4)(13+7\lambda)\over 14 c (\lambda^2-1)}
\left((\lambda-2)U_2' - 2 U_3\right)U_2 \notag \ .
\eea
(The redefinition of $L_6$, which is also needed but will not be written explicitly, is 
fixed up to a single coefficient.)  The fact that such a field redefinition is possible is already nontrivial.  
Now using the \w{\lambda} algebra  (\ref{com22})--(\ref{com44}), we compute 
$\{U_3 \ , U_3\}$, $\{U_3 \ , U_4\}$, and $\{U_4 \ , U_4\}$, and after fixing the final coefficient in the 
spin-6 operator, we find an exact match to (\ref{fmrcom}) including central terms. 
\smallskip

This exhibits the identity between \w{\lambda} and $\hat{\W}_\infty^{(\lambda)}$ by a brute-force change of basis, 
but in fact it follows from the connection between the asymptotic symmetry computation and the 
Drinfeld-Sokolov reduction, which is in turn related to the integrability framework used to construct 
$\hat{\W}_\infty^{(\lambda)}$.  As argued in \cite{Campoleoni:2010zq}, imposing the AdS boundary 
conditions in the asymptotic symmetry computation is equivalent to Drinfeld-Sokolov reduction of the 
current algebra $\widehat{\mbox{hs}}[\lambda]$ (the affinization of \hs{\lambda}).  
It was proven in \cite{Khesin:1994ey} that the Drinfeld-Sokolov reduction of $\hs{\lambda}\oplus \mathbb{C}$ gives $\W_{\rm{KP}}^{(\lambda)}$, 
and so eliminating the spin-1 field corresponding to $\mathbb{C}$ leads to the relation (\ref{fmrid}) found 
here..\footnote{The argument of \cite{Khesin:1994ey} is quite different
from what we have done here, and actually involves some extension of $\hs{\lambda}\oplus \mathbb{C}$
whose direct interpretation in the current context is not clear to us. We thank
the authors of \cite{Campoleoni:2010zq} for bringing
reference  \cite{Khesin:1994ey} to our attention.}
\section{The  \w{\lambda} CFT }\label{ss:CFT}

Now we want to switch gears and consider the problem from the point of view of the dual CFT. 
The above analysis suggests that the boundary CFT should have the $\W$-algebra
$\w{\lambda}$ as its symmetry. Thus the states of this CFT must fall into representations 
of this algebra. Our goal is to provide evidence that the 't~Hooft limit of the coset theory 
defining the minimal model CFT indeed satisfies this expectation, thereby providing a nontrivial 
check of the proposed duality. This check will be insensitive to the detailed structure of the 
nonlinear terms in \w{\lambda}, so it should be considered a check on the global symmetries.

Recall from \cite{Bouwknegt:1992wg,Gaberdiel:2010pz} that the $\W_N$ minimal model at level $k$ is the 
conformal field theory based on the coset
\be\label{coset}
\frac{{\mathfrak su}(N)_k \oplus {\mathfrak su}(N)_1}{{\mathfrak su}(N)_{k+1}} \ .
\ee
The central charge of this CFT is 
\be\label{cdef}
c = (N-1) \Bigl(1 - \frac{N}{N+k} \frac{N+1}{N+1+k} \Bigr) \ ,
\ee
and the representations that survive in the 't~Hooft limit can all be obtained by taking successive
tensor powers of the representations labelled by 
\be\label{prim}
(0;{\rm f})\ , \qquad (0,\bar{\rm f}) \ , \qquad
({\rm f};0) \ , \qquad (\bar{\rm f};0) \ .
\ee
Here $(\rho;\nu)$ labels the representation of the coset model (\ref{coset}) with 
$\rho$ being the representation of ${\mathfrak su}(N)_k$, while $\nu$ is the 
representation of ${\mathfrak su}(N)_{k +1}$. ${\rm f}$ denotes the fundamental
representation of  ${\mathfrak su}(N)$, while $\bar{\rm f}$ is the anti-fundamental
representation,  see \cite{Gaberdiel:2010pz} for
further details. In the 't~Hooft limit, the conformal dimensions of the corresponding
primary states are
\be
h(0;{\rm f}) = h(0,\bar{\rm f}) = \tfrac{1}{2} (1-\lambda) \ , \qquad
h({\rm f};0) =h(\bar{\rm f};0)= \tfrac{1}{2} (1+\lambda) \ . 
\ee
Furthermore, it was argued that their characters are of the form
\be\label{ch1}
\chi_{(0;{\rm f})}(q) = \chi_{(0;\bar{\rm f})}(q) = 
q^{\frac{1}{2}(1-\lambda) - \frac{c}{24}} \, 
\frac{1}{(1-q)}\, \prod_{s=2}^{\infty} \prod_{n=s}^{\infty} \frac{1}{(1-q^n)} \ ,
\ee
and
\be\label{ch2}
\chi_{({\rm f};0)}(q) = \chi_{(\bar{\rm f};0)}(q) = 
q^{\frac{1}{2}(1+\lambda) - \frac{c}{24}} \, 
\frac{1}{(1-q)}\, \prod_{s=2}^{\infty} \prod_{n=s}^{\infty} \frac{1}{(1-q^n)} \ .
\ee
In the following we want to show that at least these four representations 
are indeed representations of $\w{\lambda}$. 

\subsection{The Wedge Algebra}

In the 't~Hooft limit, $N\rightarrow \infty$ for fixed $\lambda$, and hence $c\rightarrow \infty$, see
(\ref{cdef}). As was argued above, the subalgebra generated by $W^s_n$ with $|n|<s$ then defines
a closed subalgebra, namely the wedge algebra. Furthermore, for the case of $\w{\lambda}$,
the wedge algebra is precisely equal to \hs{\lambda}.

Now suppose $\phi$ defines a primary state of the $\w{\lambda}$ algebra. Then it must, in particular,
define a representation of \hs{\lambda}. Conversely, any representation of \hs{\lambda} gives rise to a 
representation of $\w{\lambda}$ by the usual Verma module construction. This is to say, we postulate that
$\phi$ is annihilated by all positive modes, and define the Verma module to be the representation of
$\w{\lambda}$ that is generated by the action of the negative modes from $\phi$. This is uniquely
determined once we know the action of all zero modes on $\phi$. For specific choices of
the central charge the resulting Verma module may be reducible, but generically this representation
of  $\w{\lambda}$ will be irreducible.

The above statement may sound a bit abstract, but is familiar from many examples.
For instance, for the case of a WZW model based on $\hat{\mathfrak g}$,
this is just the statement that the representation of $\hat{\mathfrak g}$
is uniquely characterized by a representation of the finite dimensional Lie algebra ${\mathfrak g}$
on the highest weight states. The resulting representation generically does not have any null-vectors;
they only arise if $k$ is a positive integer or an admissible fractional level. Similarly, a 
Virasoro highest weight representation is uniquely characterized by the conformal dimension. 
The only generic null vector appears for $h=0$ since then $L_{-1}\phi = 0$. This null-vector
is already visible within the wedge algebra. Apart from that, at fixed $h$ there are only
additional null vectors for specific values of the central charge $c$.
\smallskip

Returning to the case at hand, we have a very explicit description of \hs{\lambda} as a quotient of the universal
enveloping algebra  of sl$(2)$, see (\ref{idealcon}), and we can hence study its representation theory 
directly. In particular, there is one simple class of representations of \hs{\lambda}: these are 
the representations of sl$(2)$ for which the quadratic Casimir takes the value $\tfrac{1}{4}(\lambda^2-1)$! 
On a highest weight state, {\it i.e.}\ a state with $J_+\phi=0$, with conformal dimension $h$, {\it i.e.}\ 
$J_0\phi=h\phi$,  the quadratic Casimir  (\ref{cas}) takes the eigenvalue
\be
C_2 \phi = h \, (h-1) \phi \ .
\ee
Thus $\phi$ has $C_2=\tfrac{1}{4}(\lambda^2-1)$ 
if $h=h_\pm=\tfrac{1}{2}(1\pm \lambda)$. Let us call
the corresponding states $\phi_\pm$, {\it i.e.}\ 
\be
J_+ \phi_\pm = 0 \ , \qquad J_0 \phi_\pm = h_\pm \phi_\pm  \ , \quad \hbox{with}
\quad h_\pm = \tfrac{1}{2}(1\pm \lambda) \ .
\ee
Both $\phi_\pm$ generate a representation of sl$(2$), that defines a representation of \hs{\lambda}. 

These two representations of \hs{\lambda} now correspond to the two representations that appeared
 in \cite{Gaberdiel:2010pz}
\be\label{pair1}
\phi_- \leftrightarrow (0;{\rm f})  \qquad \hbox{and} \qquad \phi_+ \leftrightarrow (\bar{\rm f};0)  \ .
\ee
(The reason why we group together this pair of representations will become clear momentarily.)
These are not the only representations though. The wedge algebra \hs{\lambda} has the automorphism 
\begin{equation}
V^s_n \mapsto (-1)^s V^s_n \ ,
\end{equation}
as follows immediately from the structure of the commutators in (\ref{vcom}). 
This automorphism corresponds to `charge conjugation'. 
As we shall see momentarily $V^3_0$ has a non-trivial eigenvalue 
on the above representations labelled by $h_\pm$; thus the representations corresponding
to $\phi_\pm$ are not self-conjugate, and we need to introduce their conjugate representations 
$\bar\phi_\pm$ as well. This is mirrored by the fact that also
the representations in (\ref{pair1}) are not self-conjugate; their conjugate representations are given
by the other two representations that appeared in \cite{Gaberdiel:2010pz}
\be\label{pair2}
\bar\phi_- \leftrightarrow (0;\bar{\rm f})  \qquad \hbox{and} \qquad
\bar\phi_+ \leftrightarrow  ({\rm f};0)  \ .
\ee

\subsection{Characters}

There is one simple consistency check one can immediately perform. Since the representations
$\phi_\pm$ and $\bar\phi_\pm$ are actually representations of sl$(2)$, we know the characters
of their \hs{\lambda} representations explicitly; it is simply given by
\be
\chi_{\phi_\pm}(q) = \chi_{\bar\phi_{\pm}}(q) = \frac{q^{h_\pm}}{(1-q)} \ .
\ee
The associated representation of \w{\lambda} is then simply obtained by multiplying with the
Verma module partition function coming from the negative modes that are `outside' the wedge,
{\it i.e.}\ the modes $W^s_n$ with $n\leq -s$. Thus the corresponding \w{\lambda} characters 
$\hat\chi_{\phi_{\pm}}(q)$ are then
\be
\hat\chi_{\phi_{\pm}}(q) =  \hat\chi_{\bar\phi_{\pm}}(q)  = q^{-\frac{c}{24}}\, \chi_{\phi_\pm}(q) \times 
\prod_{s=2}^{\infty} \prod_{n=s}^{\infty}  \frac{1}{(1-q^n)} \ .
\ee
This then reproduces precisely (\ref{ch1}) and (\ref{ch2}).

\subsection{The Spin 3 Zero Mode}

For the above argument it was important that the eigenvalue of $V^3_0$ does not vanish on
the highest weight state. In fact, given (\ref{V3}) it is easy to determine $V^3_0$ explicitly on 
$\phi_\pm$,
\be\label{V30}
V^3_0 \phi_\pm = \tfrac{1}{3} h_{\pm}(2h_{\pm} + 1 ) \, \phi_\pm = 
\tfrac{1}{6} (1 \pm \lambda) (2\pm \lambda) \, \phi_\pm \ .
\ee
Unless $\lambda=1$ or $\lambda=2$ this does not vanish.

Actually, we can test the above identification further by comparing these eigenvalues with
the eigenvalue of the spin $3$ mode of the coset algebra. (This will then also allow us to explain
why the representations should be paired up as in (\ref{pair1}) and (\ref{pair2}).) In the coset
description the spin $3$ field is the singlet in ${\mathfrak su}(N)_k\oplus {\mathfrak su}(N)_1$
at conformal weight three that is primary with respect to the diagonal ${\mathfrak su}(N)_{k+1}$ 
algebra. Let us denote the modes of ${\mathfrak su}(N)_k$ by $K^a_n$, while those of
${\mathfrak su}(N)_1$ will be denoted by $J^a_n$. We make the ansatz for the singlet state
at conformal weight three to be of the form
\be\label{Wans}
W = d_{abc} \Bigl( a_1 K^a_{-1} K^b_{-1} K^c_{-1} 
+ a_2 K^a_{-1} K^b_{-1} J^c_{-1} 
+ a_3 K^a_{-1} J^b_{-1} J^c_{-1} 
+ a_4 J^a_{-1} J^b_{-1} J^c_{-1}  \Bigr) \Omega \ ,
\ee
where $d_{abc}$ is the (unique) symmetric traceless invariant tensor of rank $3$ for 
sl$(N)$ (with $N\geq 3$). The condition that $W$ is primary with respect to ${\mathfrak su}(N)_{k+1}$
means that is must be annihilated by $K^d_1+J^d_1$ for all $d$. Using the commutation relations
\be
{} [K^a_m, K^b_n] = f_{abc} K^c_{m+n} + k \, m \delta^{ab} \delta_{m,-n} \ , \qquad
{} [J^a_m, J^b_n] = f_{abc} J^c_{m+n} + m \delta^{ab} \delta_{m,-n} \ ,
\ee
this leads to the relations (see \cite{Bais:1987zk})
\be
3 (k+N) a_1+ a_2 = 0 \ , \qquad
(2k+N) a_2 + (2+N) a_3 = 0 \ , \qquad
k a_3 + 3 (1+N) a_4 = 0 \ ,
\ee
where we have used the tensor identity (see for example \cite[Appendix B]{Bais:1987dc})
\be
d_{abc} f_{dae} f_{ebg} = N d_{cdg} \ .
\ee
This determines the state uniquely, up to an overall normalization. In the 
't~Hooft limit we get (note that the term proportional to $K\cdot K \cdot K$ drops out in this limit)
\be\label{WtHooft}
W = d_{abc} \Bigl( 3 \frac{\lambda^2}{(1-\lambda) (2-\lambda)} \, K^a_{-1} K^b_{-1} J^c_{-1} 
- 3 \frac{\lambda}{(1-\lambda)} \, K^a_{-1} J^b_{-1} J^c_{-1}  + 
J^a_{-1} J^b_{-1} J^c_{-1}  \Bigr) \Omega \ .
\ee
Now we can evaluate the zero mode of this state on the primary states (\ref{prim}). For
the states $(0;{\rm f})$ and $(0;\bar{\rm f})$ this is straightforward since $K^a_0=0$, and hence
we simply get
\be\label{W1}
W_0 (0;{\rm f}) = C (0;{\rm f}) \ , \qquad 
W_0 (0;\bar{\rm f}) = - C (0;\bar{\rm f})  \ ,
\ee
where $C$ is an (unimportant) constant defined by 
\be
d_{abc} J^a_0 J^b_0 J^c_0 \, |{\rm f}\rangle = C \, |{\rm f}\rangle \ , \qquad
d_{abc} J^a_0 J^b_0 J^c_0 \, |\bar{\rm f}\rangle = - C \, |\bar{\rm f}\rangle \ .
\ee
On the other hand, for the states $({\rm f};0)$ and $(\bar{\rm f};0)$ the analysis is more subtle. 
In the first case, the ground states transform as $[{\rm f},\bar{\rm f}]$ with respect to $K^a_0$, $J^a_0$, 
but since we are only interested in the singlet component with respect to the diagonal,
we have $K^a_0+J^a_0=0$. (Similarly, in the second case, the ground states 
transform as $[\bar{\rm f},{\rm f}]$
with respect to $K^a_0$, $J^a_0$, and we are again only interested in the singlet component,
{\it i.e.}\ the linear combination that is annihilated by $K^a_0+J^a_0$.) Using the singlet condition to replace
$K^a_0$ by $J^a_0$ we then get
\bea\label{W2}
W_0 ({\rm f};0) & = & \Bigl( 3 \frac{\lambda^2}{(1-\lambda) (2-\lambda)}  
+ 3 \frac{\lambda}{(1-\lambda)}  + 1 \Bigr) d_{abc} J^a_{0} J^b_{0} J^c_{0}\,  ({\rm f};0)  \notag \\
& = & - C\, \frac{(1+\lambda) (2+\lambda)}{(1-\lambda) (2-\lambda)} \, ({\rm f};0)  \ ,
\eea
and similarly
\be\label{W2p}
W_0 (\bar{\rm f};0) =  C\, \frac{(1+\lambda) (2+\lambda)}{(1-\lambda) (2-\lambda)} \, (\bar{\rm f};0) \ .
\ee

Now we can compare these results with the action of $V^3_0$ on the primary states
$\phi_\pm$ and $\bar\phi_\pm$. A priori, we do not know how to fix the relative normalisation between
$V^3_0$ and $W_0$. However, if we want to identify $\phi_- \leftrightarrow (0;{\rm f})$, see
eq.\ (\ref{pair1}), it follows that we must have
\be\label{norma}
V^3_0 = \frac{(1-\lambda) (2-\lambda)}{C} W_0 \ .
\ee
Having fixed the relative normalisation, we can now compare the eigenvalues on the remaining
states. In particular we find, using (\ref{norma}) as well as (\ref{W1}) and (\ref{W2}), (\ref{W2p})
\bea
V^3_0 (0;\bar{\rm f}) &  = &  - (1-\lambda) (2-\lambda)\, (0;\bar{\rm f}) \\
V^3_0 ({\rm f};0) & = & - (1+\lambda) (2+\lambda)\, ({\rm f};0) \\
V^3_0 (\bar{\rm f};0) & = &  (1+\lambda) (2+\lambda)\, (\bar{\rm f};0)  \ .
\eea
Given (\ref{V30}) this is then in perfect agreement with the identifications (\ref{pair1}) and (\ref{pair2}). 

One may wonder whether one could repeat the analysis for the eigenvalue of the spin $4$ 
field, but it is not clear to us how to do this. On the \hs{\lambda} representations it is again straightforward to 
calculate the eigenvalues of $V^4_0$. This mode should now be identified with a zero mode of a spin
$4$ state in the coset theory. However, at conformal dimension four, there is the analogue of
(\ref{Wans}), but also the quasiprimary state associated to the normal ordered product of $:LL:$. 
In relating the mode $V^4_0$ with the zero mode of the spin 4 state, there are then two unknown 
parameters (namely the coefficient in front of the analogue of (\ref{Wans}), and the coefficient
in front of $:LL:$), and we cannot make any check, unless there is some independent way of fixing the 
normalizations.

\bigskip

\noindent
{\bf Acknowledgements:} We thank Jos\'{e} Figueroa-O'Farrill, Liam Fitzpatrick, Daniel Jafferis, 
Rajesh Gopakumar, Jonathan Heckman, and Soo-Jong Rey for helpful discussions and correspondences. 
The stay of MRG at the IAS was partially supported by The Ambrose Monell Foundation. 
His work is also supported in part by the Swiss National Science Foundation.
TH is supported by U.S. Department of Energy grant DE-FG02-90ER40542.

\section*{Appendix} 
\appendix

\section{Structure Constants of Higher Spin Algebras}\label{a:structure}
The higher spin algebra \hs{\lambda} has commutators (\ref{vcom}) with structure constants \cite{Pope:1989sr}
\bea\label{struct}
g_u^{st}(m,n;\lambda) &=& \hspace*{-0.1cm}{2 q^{u-2}  \over (u-1)!} \phi_u^{st}(\lambda) N_u^{st}(m,n)\\
N_u^{st}(m,n) &=& \hspace*{-0.1cm} \sum_{k=0}^{u-1}(-1)^k
\left(\!\!\begin{array}{c}u-1 \\k\end{array}\!\!\right)
[s-1+m]_{u-1-k}[s-1-m]_k[t-1+n]_k[t-1-n]_{u-1-k}\notag\\
\phi_u^{st}(\lambda) &=&\hspace*{-0.1cm}  \ _4F_3\left[
\begin{array}{c|}
\half + \lambda \ ,  \ \half - \lambda \ , {2-u\over 2} \ , {1-u\over 2}\\
{3\over 2}-s \ , \ {3\over 2} -t\ , \ \half + s+t-u
\end{array}  \ 1\right]\notag \ ,
\eea
where $ [a]_n \equiv \Gamma(a+1)/\Gamma(a+1-n)$ is the descending Pochhammer symbol. 
$q$ is an arbitrary number that can be scaled to $q=1$, but it is useful to keep explicitly because 
$q$ accounts for all possible rescalings of the generators consistent with the leading term 
in the commutator as well as the usual normalization 
of the sl$(2)$ subalgebra and its action on higher spin generators (\ref{twoaction}).  In the 
enveloping algebra construction (\ref{env}), $q = {1\over 4}$, whereas in the discussion of 
the $\lambda \to \infty$ limit we scaled $q \sim 1/\lambda$.  In the comparison to  
$\hat{\W}_{\infty}^{(\lambda)}$ in section \ref{ss:fmr} we have also set $q={1\over 4}$.

A few special values of the structure constants are useful.  For $\lambda=\half$, we have 
$\phi^{st}_u(\half) = 1$ and the algebra becomes hs$(1,1)$.  In the asymptotic symmetry 
computation for general $\lambda$, only structure constants with $m=-s+1$ appear in (\ref{coV}), 
and these simplify to
\be\label{specialg}
g^{st}_u(-s+1,n;\lambda) = \frac{ (-1)^{u+1} q^{u-2}\Gamma(2s-1)
\Gamma(n+t)}{2\Gamma(2s-u)\Gamma(1+n+t-u)\Gamma(u)}\, \phi^{st}_u(\lambda) \ .
\ee
The quadratic form (\ref{quadform}) is explicitly
\bea\label{appquadform}
\tr(V^s_m V^t_n) &\equiv& {3\over 4 q (\lambda^2-1)} g^{s t}_{s+t-1}(m,n,\lambda)\\
&=&N_s  {(-1)^{s-m-1}\over (2s-2)!}\Gamma(s+m)\Gamma(s-m)  \delta^{st}\delta_{m,-n} \notag\\
N_s &\equiv& {3 \cdot 4^{s-3}\sqrt{\pi}q^{2s-4}\Gamma(s)\over (\lambda^2-1)
\Gamma(s+\half)} (1-\lambda)_{s-1} (1+\lambda)_{s-1} \ , \label{appN}
\eea
where $(a)_n = \Gamma(a+n)/\Gamma(a)$ is the ascending Pochhammer symbol 
and the overall constant has been chosen to set
\be
\tr(V^2_1 V^2_{-1}) = -1 \ .
\ee

\section{Full Commutation Relations of \w{\lambda}}\label{a:fmr}

In this appendix we reproduce the commutation relations of the algebra 
$\hat{\W}_\infty^{(\lambda)}$ \cite{FigueroaO'Farrill:1992cv} 
which after the change of 
basis described in section \ref{ss:fmr} is equivalent to \w{\lambda}. The Dirac bracket 
of currents is (\ref{fmrcom}), where $P_{st}(\phi)$ has two contributions,
\be\label{fullcom}
P_{st}(\phi) = P_{st}^{\rm KP}(\phi) + \delta P_{st}(\phi) \ .
\ee
The first contribution is the commutator of $\W_{KP}^{(\lambda)}$, including the spin-1 field,
\bea
P_{st}^{\rm KP} &=& {c\over \lambda}\sum_{u=1}^s \bif{t-\lambda-1}{t+u-1}\bif{\lambda}{s-u}
\p^{s+t-1}-\sum_{u=1}^s\bif{s-1}{u-1}U_{t+u-1}(-\p)^{s-u}\\
& & +\sum_{u=1}^{s-1}\sum_{r=1}^{s-u}\bif{t-\lambda-1}{t+u-1}\bif{\lambda-r}{s-r-u}U_r\p^{s+t-r-1} \notag\\
& & +\sum_{u=1}^s\sum_{r=1}^{t+u-1}\bif{t-\lambda-1}{t+u-r-1}\bif{\lambda}{s-u}\p^{s+t-r-1}U_r \notag\\
& & -{\lambda\over c}\sum_{u=1}^{s-1}\sum_{r=1}^{s-u}\bif{s-r-1}{u-1}U_{t+u-1}(-\p)^{s-u-r}U_r \notag\\
& & + {\lambda \over c}\sum_{u=1}^{s-1}\sum_{r=1}^{s-u}\sum_{p=1}^{u+t-1}
\bif{t-\lambda-1}{t+u-p-1}\bif{\lambda-r}{s-r-u}U_r \p^{s+t-r-p-1} U_p \ ,\notag
\eea
where $\p = \p_\phi$, and currents are evaluated at $\phi$. The second contribution, which comes from 
imposing the constraint $U_1=0$ and going to Dirac brackets, is
\bea
\delta P_{st} &=& {c\over \lambda^2}(-1)^{s-1}\bif{s-\lambda-1}{s}\bif{t-\lambda-1}{t}\p^{s+t-1}\\
& &  + {(-1)^{s-1}\over \lambda}\bif{s-\lambda-1}{s}\sum_{r=2}^{t-1}\bif{t-\lambda-1}{t-r}\p^{s+t-r-1}U_r \notag\\
& & + {(-1)^t\over \lambda}\bif{t-\lambda-1}{t}\sum_{r=2}^{s-1}\bif{s-\lambda-1}{s-r}U_r(-\p)^{s+t-r-1} \notag\\
& & + {(-1)^{s-1}\over c}\sum_{r=2}^{s-1}\sum_{u=2}^{t-1}\bif{s-\lambda-1}{s-r}
\bif{t-\lambda-1}{t-u}(-1)^r U_r \p^{s+t-r-u-1}U_u  \ .\notag
\eea
It is was conjectured in \cite{FigueroaO'Farrill:1992cv} that $\lambda$ is not a true parameter 
of the algebra $\W_{\rm KP}^{(\lambda)}$, \textit{i.e.}, that different values of $\lambda$ are simply 
different choices of basis for the same algebra (except for integer values of $\lambda$).  However,
after the reduction, $\lambda$ becomes a true parameter, and the algebras 
$\hat{\W}_\infty^{(\lambda)}$ are inequivalent for different values of $\lambda$.\footnote{Note that 
there is a typo in Conjecture 4.14 of \cite{FigueroaO'Farrill:1992cv}:  $\hat{\W}_\infty^{(\lambda)}$ should 
be replaced by $\W_{\rm KP}^{(\lambda)}$. We thank Jos\'e Figueroa-O'Farrill for clarification of this point.}


\begin{thebibliography}{99}

\bibitem{Koch:2010cy}
R.d.M.~Koch, A.~Jevicki, K.~Jin and J.P.~Rodrigues,
``AdS$_4$/CFT$_3$ construction from collective fields,''
arXiv:1008.0633 [hep-th].
  
\bibitem{Douglas:2010rc}
M.R.~Douglas, L.~Mazzucato and S.S.~Razamat,
``Holographic dual of free field theory,''
arXiv:1011.4926 [hep-th].
  
\bibitem{Coleman:1974bu}
S.R.~Coleman,
``Quantum sine-Gordon equation as the massive Thirring model,''
Phys.\ Rev.\  D {\bf 11}, 2088 (1975).

\bibitem{Klebanov:2002ja}
I.R.~Klebanov and A.M.~Polyakov,
``AdS dual of the critical O(N) vector model,''
Phys.\ Lett.\  B {\bf 550}, 213 (2002)
[arXiv:hep-th/0210114].
  
\bibitem{Vasiliev:2003ev}
M.A.~Vasiliev,
``Nonlinear equations for symmetric massless higher spin fields in (A)dS(d),''
Phys.\ Lett.\  B {\bf 567}, 139 (2003)
[arXiv:hep-th/0304049].

\bibitem{Bekaert:2005vh}
X.~Bekaert, S.~Cnockaert, C.~Iazeolla and M.A.~Vasiliev,
``Nonlinear higher spin theories in various dimensions,''
arXiv:hep-th/0503128.
  
\bibitem{Giombi:2009wh}
S.~Giombi and X.~Yin,
``Higher spin gauge theory and holography: the three-point functions,''
JHEP {\bf 1009}, 115 (2010)
[arXiv:0912.3462 [hep-th]].

\bibitem{Giombi:2010vg}
S.~Giombi and X.~Yin,
``Higher spins in AdS and twistorial holography,''
  arXiv:1004.3736 [hep-th].
  
\bibitem{Brown:1986nw}
J.D.~Brown and M.~Henneaux,
``Central charges in the canonical realization of asymptotic symmetries: an
example from three-dimensional gravity,''
Commun.\ Math.\ Phys.\  {\bf 104}, 207 (1986).

\bibitem{Kiritsis:2010xc}
E.~Kiritsis and V.~Niarchos,
``Large-N limits of 2d CFTs, quivers and AdS$_3$ duals,''
arXiv:1011.5900 [hep-th].

\bibitem{Henneaux:2010xg} 
M.~Henneaux and S.-J.~Rey,
``Nonlinear $W_{\infty}$ as asymptotic symmetry of three-dimensional higher spin 
Anti-de Sitter gravity,''
JHEP {\bf 1012}, 007 (2010)
[arXiv:1008.4579 [hep-th]].

\bibitem{Campoleoni:2010zq}
A.~Campoleoni, S.~Fredenhagen, S.~Pfenninger and S.~Theisen,
``Asymptotic symmetries of three-dimensional gravity coupled to higher-spin fields,''
JHEP {\bf 1011}, 007 (2010)
[arXiv:1008.4744 [hep-th]].

\bibitem{Gaberdiel:2010ar}
M.R.~Gaberdiel, R.~Gopakumar and A.~Saha,
``Quantum W-symmetry in AdS$_3$,''
arXiv:1009.6087 [hep-th].

\bibitem{Gaberdiel:2010pz}
M.R.~Gaberdiel and R.~Gopakumar,
``An AdS3 dual for minimal model CFTs,''
arXiv:1011.2986 [hep-th].
 
\bibitem{Strominger:1996sh}
A.~Strominger and C.~Vafa,
``Microscopic origin of the Bekenstein-Hawking entropy,''
Phys.\ Lett.\  B {\bf 379}, 99 (1996)
[arXiv:hep-th/9601029].

\bibitem{Maldacena:1997re}
J.M.~Maldacena,
``The large N limit of superconformal field theories and supergravity,''
Adv.\ Theor.\ Math.\ Phys.\  {\bf 2}, 231 (1998)
[Int.\ J.\ Theor.\ Phys.\  {\bf 38}, 1113 (1999)]
[arXiv:hep-th/9711200].
  
\bibitem{Castro:2010ce}
A.~Castro, A.~Lepage-Jutier and A.~Maloney,
``Higher spin theories in AdS$_3$ and a gravitational exclusion principle,''
arXiv:1012.0598 [hep-th].

\bibitem{FigueroaO'Farrill:1992cv}
J.M.~Figueroa-O'Farrill, J.~Mas and E.~Ramos,
``A one parameter family of Hamiltonian structures for the KP hierarchy and a
continuous deformation of the nonlinear W(KP) algebra,''
Commun.\ Math.\ Phys.\  {\bf 158}, 17 (1993)
[arXiv:hep-th/9207092].
  
\bibitem{Khesin:1993ru} 
 B.~Khesin and I.~Zakharevich,
 ``Poisson - Lie group of pseudodifferential symbols,''
Commun.\ Math.\ Phys.\  {\bf 171}, 475 (1995)
[arXiv:hep-th/9312088].

\bibitem{Khesin:1993ww} 
B.~Khesin and I.~Zakharevich,
``Poisson Lie group of pseudodifferential symbols and fractional KP - KdV hierarchies,''
arXiv:hep-th/9311125.
  
\bibitem{Pope:1989ew}
C.N.~Pope, L.J.~Romans, X.~Shen,
``The complete structure of W($\infty$),''
Phys.\ Lett.\  B {\bf 236}, 173 (1990).

\bibitem{Bowcock:1991zk}
P.~Bowcock and G.M.T.~Watts,
``On the classification of quantum W algebras,''
Nucl.\ Phys.\  B {\bf 379}, 63 (1992)
[arXiv:hep-th/9111062].

 \bibitem{workinprog}
M.R.~Gaberdiel, R.~Gopakumar, T.~Hartman, and S.~Raju, {\it work in progress}.
  
\bibitem{Pope:1989sr}
C.N.~Pope, L.J.~Romans and X.~Shen,
``W($\infty$) and the Racah-Wigner algebra,"
Nucl.\ Phys.\  B {\bf 339}, 191 (1990).

\bibitem{Feigin88} 
B.L.~Feigin, 
``Lie algebras gl$(\lambda)$ and cohomology of a Lie algebra of differential operators,"
Russian Mathematical Surveys {\bf 43} no.\ 2, 169  (1988).

\bibitem{Bordemann:1989zi}
M.~Bordemann, J.~Hoppe and P.~Schaller,
``Infinite dimensional matrix algebras,"
Phys.\ Lett.\  B {\bf 232}, 199 (1989).

\bibitem{Bergshoeff:1989ns}
E.~Bergshoeff, M.P.~Blencowe and K.S.~Stelle,
``Area preserving diffeomorphisms and higher spin algebra,''
Commun.\ Math.\ Phys.\  {\bf 128}, 213 (1990).

\bibitem{Vasiliev:1989re}
M.A.~Vasiliev,
``Higher spin algebras and quantization on the sphere and hyperboloid,"
Int.\ J.\ Mod.\ Phys.\  A {\bf 6}, 1115 (1991).
  
\bibitem{Fradkin:1990qk}
E.S.~Fradkin and V.Y.~Linetsky,
``Supersymmetric Racah basis, family of infinite dimensional superalgebras, SU($\infty$ + 1$|\infty$) and 
related 2-D models,''
Mod.\ Phys.\ Lett.\  A {\bf 6}, 617 (1991).

\bibitem{Pope:1990rn}
C.N.~Pope, L.J.~Romans and X.~Shen,
``A brief history of W($\infty$),''
talk given at Strings '90.

\bibitem{Fradkin:1986ka}
E.S.~Fradkin, M.A.~Vasiliev,
``Candidate to the role of higher spin symmetry,''
Annals Phys.\  {\bf 177}, 63 (1987).
  
\bibitem{Blencowe:1988gj}
M.P.~Blencowe,
``A consistent interacting massless higher spin field theory in $D = (2+1)$,''
Class.\ Quant.\ Grav.\  {\bf 6}, 443 (1989).
  
\bibitem{Pope:1989cr}
C.N.~Pope and K.S.~Stelle,
``SU($\infty$), SU+($\infty$) and area preserving algebras,"
Phys.\ Lett.\  B {\bf 226}, 257 (1989).

\bibitem{Zamolodchikov:1985wn}
A.B.~Zamolodchikov,
``Infinite additional symmetries in two-dimensional conformal quantum field theory,''
Theor.\ Math.\ Phys.\  {\bf 65}, 1205 (1985).

\bibitem{Drinfeld:1984qv}
V.G.~Drinfeld and V.V.~Sokolov,
``Lie algebras and equations of Korteweg-de Vries type,''
J.\ Sov.\ Math.\  {\bf 30}, 1975 (1984).

\bibitem{Balog:1990dq}
J.~Balog, L.~Feher, P.~Forgacs, L.~O'Raifeartaigh and A.~Wipf,
``Kac-Moody realization of W algebras,"
Phys.\ Lett.\  B {\bf 244}, 435 (1990).
  
\bibitem{Bouwknegt:1992wg}
P.~Bouwknegt and K.~Schoutens,
``W symmetry in conformal field theory,''
Phys.\ Rept.\  {\bf 223}, 183 (1993)
[arXiv:hep-th/9210010].

\bibitem{Regge:1974zd}
T.~Regge and C.~Teitelboim,
``Role of surface integrals in the Hamiltonian formulation of general relativity,''
Annals Phys.\  {\bf 88}, 286 (1974).

\bibitem{Coussaert:1995zp}
O.~Coussaert, M.~Henneaux and P.~van Driel,
``The Asymptotic dynamics of three-dimensional Einstein gravity with a
negative cosmological constant,''
Class.\ Quant.\ Grav.\  {\bf 12}, 2961 (1995)
[arXiv:gr-qc/9506019].

\bibitem{Banados:1998pi}
 M.~Banados, K.~Bautier, O.~Coussaert, M.~Henneaux and M.~Ortiz,
 ``Anti-de Sitter/CFT correspondence in three-dimensional supergravity,''
Phys.\ Rev.\  D {\bf 58}, 085020 (1998)
 [arXiv:hep-th/9805165].

\bibitem{Henneaux:1999ib}
M.~Henneaux, L.~Maoz and A.~Schwimmer,
``Asymptotic dynamics and asymptotic symmetries of three-dimensional
extended AdS supergravity,''
Annals Phys.\  {\bf 282}, 31 (2000)
[arXiv:hep-th/9910013].

\bibitem{Henneaux:2002wm}
 M.~Henneaux, C.~Martinez, R.~Troncoso and J.~Zanelli,
 ``Black holes and asymptotics of 2+1 gravity coupled to a scalar field,''
Phys.\ Rev.\  D {\bf 65}, 104007 (2002)
[arXiv:hep-th/0201170].
  
 \bibitem{Henneaux:2004zi}
M.~Henneaux, C.~Martinez, R.~Troncoso and J.~Zanelli,
``Asymptotically anti-de Sitter spacetimes and scalar fields with a
logarithmic branch,''
Phys.\ Rev.\  D {\bf 70}, 044034 (2004)
[arXiv:hep-th/0404236].

\bibitem{Pope:1990kc}
C.N.~Pope, L.J.~Romans, X.~Shen,
``A new higher spin algebra and the lone star product,''
Phys.\ Lett.\  B {\bf 242}, 401 (1990).

\bibitem{Pope:1990be}
C.N.~Pope, L.J.~Romans, X.~Shen,
``Ideals of Kac-Moody algebras and realizations of W($\infty$),''
Phys.\ Lett.\  B {\bf 245}, 72 (1990).

\bibitem{Bergshoeff:1990cz}
E.~Bergshoeff, M.A.~Vasiliev and B.~de Wit,
``The SuperW($\infty$) (lambda) algebra,''
Phys.\ Lett.\  B {\bf 256}, 199 (1991).

\bibitem{Bergshoeff:1991dz}
E.~Bergshoeff, B.~de Wit and M.A.~Vasiliev,
``The structure of the superW($\infty$) (lambda) algebra,''
Nucl.\ Phys.\  B {\bf 366}, 315 (1991).

\bibitem{Bakas:1991fs}
I.~Bakas and E.~Kiritsis,
``Beyond the large N limit: Nonlinear W($\infty$) as symmetry of the Sl(2,R)/ U(1) coset model,''
Int.\ J.\ Mod.\ Phys.\  A {\bf 7}, 55 (1992)
[arXiv:hep-th/9109029].

\bibitem{Bakas:1991gs}
I.~Bakas, B.~Khesin and E.~Kiritsis,
``The logarithm of the derivative operator and higher spin algebras of
W-infinity type,''
Commun.\ Math.\ Phys.\  {\bf 151}, 233 (1993).
  
\bibitem{Yamagishi:1991ax}
K.~Yamagishi,
``A Hamiltonian structure of KP hierarchy, W (1+infinity) algebra and selfdual gravity,''
Phys.\ Lett.\  B {\bf 259}, 436 (1991).
  
\bibitem{FigueroaO'Farrill:1991ek}
J.M.~Figueroa-O'Farrill, J.~Mas and E.~Ramos,
``Bihamiltonian structure of the KP hierarchy and the W(KP) algebra,''
Phys.\ Lett.\  B {\bf 266}, 298 (1991).

\bibitem{Yu:1991ng}
F.~Yu and Y.S.~Wu,
``Hamiltonian structure, (anti)selfadjoint flows in KP hierarchy and the
W(1+infinity) and W(infinity) algebras,''
Phys.\ Lett.\  B {\bf 263}, 220 (1991).
  
\bibitem{Yu:1991bk}
F.~Yu and Y.~S.~Wu,
``Nonlinearly deformed W(infinity) algebra and second Hamiltonian structure
of KP hierarchy,''
Nucl.\ Phys.\  B {\bf 373}, 713 (1992).

\bibitem{Khesin:1994ey} 
B.~Khesin and F.~Malikov,
``Universal Drinfeld-Sokolov reduction and matrices of complex size,''
Commun.\ Math.\ Phys.\  {\bf 175}, 113 (1996)
[arXiv;hep-th/9405116].

\bibitem{Bais:1987zk}
F.A.~Bais, P.~Bouwknegt, M.~Surridge and K.~Schoutens,
``Coset construction for extended Virasoro algebras,''
Nucl.\ Phys.\  B {\bf 304}, 371 (1988).

\bibitem{Bais:1987dc}
F.A.~Bais, P.~Bouwknegt, M.~Surridge and K.~Schoutens,
``Extensions of the Virasoro algebra constructed from Kac-Moody algebras 
using higher order Casimir invariants,''
Nucl.\ Phys.\  B {\bf 304}, 348 (1988).




\end{thebibliography}
\end{document}